\documentclass[a4paper,10pt]{article}
\usepackage[dvips]{graphicx}
\usepackage{amssymb,amsmath}
\numberwithin{equation}{section}

\begin{document}

\title{ F(T)  gravity and k-essence}
\author{Ratbay Myrzakulov\footnote{Email: rmyrzakulov@gmail.com; rmyrzakulov@csufresno.edu} \\ \textit{Eurasian International Center for Theoretical Physics and  Department of General } \\ \textit{ $\&$  Theoretical Physics, Eurasian National University, Astana 010008, Kazakhstan}}

%\date{}

\date{}
\maketitle
\begin{abstract}
Modified teleparallel gravity theory with the torsion scalar have recently gained a lot of attention as a possible explanation of dark energy. We perform a thorough  reconstruction analysis on the so-called $F(T)$ models, where $F(T)$ is some general function of the torsion term,  and derive conditions for the equivalence between of $F(T)$ models  with purely kinetic k-essence. We present a new class   models of $F(T)$-gravity and  k-essence. We also proposed some new models of generalized gases and knot universes as well as some generalizations of $F(T)$ gravity. 
\end{abstract}
\vspace{2cm} 

\sloppy

%\tableofcontents
\section{Introduction}
The discovery of the accelerated expansion of the universe \cite{Perl} has forced a profound shift in our cosmological paradigm. This discovery  indicates that the universe is very nearly spatially flat and consists of about 70\% of dark energy (DE) which drives the cosmic acceleration.  The equation of state (EoS) parameter $w$ for DE should be $w<-1/3$ to maintain this acceleration.  The modern constraints on the EoS parameter are around $w=-1$. Here we can note that from the theoretical point of view there are three essentially different cases: $w>-1$ (quintessence), $w=-1$ (cosmological constant), and $w<-1$ (phantom).  The cosmological constant can explain the present accelerated expansion of the universe, for which $w=-1$. Although, the cosmological constant  is the simplest  candidate for DE, but there are serious theoretical problems associated with it such as the fine-tuning problem, the coincidence problem and so on.    To solve the cosmological constant problems, some scalar-field models (phantom fields, k-essence and so on) are proposed. These scalar-field models of inflation and dark energy correspond to a modification of the energy momentum tensor in Einstein equations. 

The other alternative approach dealing with the acceleration problem of the universe is changing the gravity law through the modification of action of gravity by means of using $F(R), F(G)$ and $F(R,G)$  instead of the Einstein-Hilbert action (see, e.g. recent  reviews \cite{N1}-\cite{CST}). Here the Lagrangian density of modified gravity theories $F$ is an arbitrary function of $R$, $G$ or both $R$ and $G$. The field equations of these modified gravity theories are 4th order that making it difficult to obtain both exact and numerical solutions. Recent very updated review of modified gravity is given in 
\cite{14}-\cite{15}. In fact, it is demonstrated that alternative gravity, especially $f(R)$ one, may give very realistic  values of $w_{DE}$ and other cosmographic parameters being very close to values of LCDM. Hence, alternative gravity may be viable candidate for effective LCDM or k-essence late-time acceleration. The technically complicated property of modified gravity is the fact that equations of motion are of higher order.

Recently, however, some models with the field equations of 2nd order [so-called $F(T)$-gravity]  are proposed \cite{BF}-\cite{L3}. These models based on the "teleparallel" equivalent of General Relativity (TEGR) \cite{Eins}-\cite{FF2}, which, instead of using the curvature defined via the Levi-Civita connection, uses the Weitzenb$\ddot{o}$ck connection that has no curvature but only torsion. In \cite{FF1}-\cite{FF2}, some models based on  modified teleparallel gravity  were presented as an alternative to inflationary models. The fact that the field equations of $F(T)$ gravity are always 2nd order makes these theories simpler than the  other modified gravity theories like $F(R)$ or $F(G)$.   More recently, some properties of $F(T)$ gravity were studied in \cite{M6}-\cite{M7}. For instance, it is demonstrated recently in \cite{36} that $f(T)$ gravity may have very realistic cosmographic parameters, fitting it with liminocity distance and BAO. In fact, $f(T)$ cosmography \cite{36} may give the regions overlapping with LCDM model. It is clear that $F(T)$ gravity presents a very rich behavior and deserves further investigation.

The purpose of  the present paper is to investigate  some models of $F(T)$ gravity as well as k-essence. Also we will study the equivalence of modified gravity theories with k-essence.

This paper is organized as follows. In the  following section we review $F(T)$ gravity and present some its models.  In Sec. III we investigate some models of k-essence. The relation between $F(T)$ gravity and k-essence is studied in Sec.IV. In the last section we will give some conclusions.

\section{$F(T)$ gravity}
\subsection{Elements  of $F(T)$  gravity}

The action of   $F(T)$ - gravity   reads as  (see, e.g. \cite{BF}, \cite{L3}, \cite{M6})
\begin {equation}
S=\int d^{4}xe[\frac{1}{2\kappa^{2}}F(T)+L_{m}],
\end{equation} 
where $T$ is the torsion scalar, $e=\det{(e^{i}_{\mu})}=\sqrt{-g}$ and $L_m$ stands for the matter Lagrangian. Here $e^{i}_{\mu}$ are the components of the vierbein vector field $\textbf{e}_{A}$ in a coordinate basis, that is $\textbf{e}_{A}\equiv e^{\mu}_{A}\partial_{\mu}$. Note that in the teleparallel gravity, the dynamical variable is the vierbein field $\textbf{e}_{A}(x^{\mu})$.  The variation of the action with respect to this vierbein field leads to  the following gravitational equations of motion
\begin {equation}
[e^{-1}\partial_{\mu}(eS^{\mu\nu}_{i})-e^{\lambda}_{i}T^{\rho}_{\mu\lambda}S^{\nu\mu}_{\rho}]F_{T}+S^{\mu\nu}_{i}(\partial_{\mu}T)F_{TT}+\frac{1}{4}e^{\nu}_{i}F=\frac{1}{2}k^2e^{\rho}_{i}T^{\nu}_{\rho}.
\end{equation}
Here the torsion scalar  
$T$  is given by
\begin{equation}
T=S^{\mu\nu}_{\rho}T^{\rho}_{\mu\nu}
\end{equation}
with
\begin{equation}
	S_\rho\,^{\mu\nu}=\frac{1}{2}(K^{\mu\nu}\,_\rho+\delta^\mu_\rho T^{\theta\nu}\,_\theta-\delta^\nu_\rho T^{\theta\mu}\,_\theta).
\end{equation}
Here the contorsion tensor is defined as
\begin{equation}
K^{\mu\nu}\,_\rho=-\frac{1}{2}(T^{\mu\nu}\,_\rho-T^{\nu\mu}\,_\rho-T_\rho\,^{\mu\nu})
\end{equation}
and the torsion tensor looks like
\begin{equation}
T^\lambda_{\mu\nu}=\stackrel{w}{\Gamma}^\lambda_{\nu\mu}-\stackrel{w}{\Gamma}^\lambda_{\mu\nu}=e^\lambda_i(\partial_\mu e^i_\nu-\partial_\nu e^i_\mu).
\end{equation}
The vierbein vector fields relate with the metric through
\begin{equation}
g_{\mu\nu}(x)=\eta_{ij}e^i_\mu(x)e^j_\nu(x), 
\end{equation}
where $ {\bf e}_i \cdot {\bf e}_j=\eta_{ij}$ and $\eta_{ij}=diag(1, -1, -1, -1).$
We now will assume a flat homogeneous and isotropic FRW universe with  the metric 
\begin{equation}
ds^{2}=-dt^{2}+a(t)^{2}\sum^{3}_{i=1}(dx^{i})^{2},
\end{equation}
where $t$ is  cosmic time. Then the modified Friedmann equations and the continuity equation read as (see, e.g. \cite{BF}, \cite{L3}, \cite{M6})
\begin{equation}
	-2TF_{T}+F=2k^2 \rho_m, 
\end{equation}
\begin{equation}
	-8\dot{H}TF_{TT}+(2T-4\dot{H})F_{T}-F=2k^2p_m, 
\end{equation}
\begin{equation}
	\dot{\rho}_m+3H(\rho_m+p_m)=0. 
\end{equation}
This set  can be rewritten as
\begin{equation}
	-T-2Tf_{T}+f=2k^2 \rho_m, 
\end{equation}
\begin{equation}
	-8\dot{H}Tf_{TT}+(2T-4\dot{H})(1+f_{T})-T-f=2k^2p_m, 
\end{equation}
\begin{equation}
	\dot{\rho}_m+3H(\rho_m+p_m)=0, 
\end{equation}
with the action
\begin {equation}
S=\int d^{4}xe[\frac{1}{2\kappa^{2}}(T+f(T))+L_{m}],
\end{equation} where $f=F-T.$  Some properties of $F(T)$ - gravity  were studied in \cite{L3}-\cite{M7}. 
Note that we can rewrite the gravitational equations (2.9)-(2.10) as
\begin{equation}
	\hat{M}_1F=2k^2 \rho_m, 
\end{equation}
\begin{equation}
\hat{M}_2F=-\hat{M}_{3}\hat{M}_{1}F=2k^2p_m, 
\end{equation}
\begin{equation}
\hat{M}_3\rho_{m}=-p_{m}, 
\end{equation}
where
\begin{equation}
\hat{M}_1=	-2T\partial_{T}+1, 
\end{equation}
\begin{equation}
\hat{M}_2=	-8\dot{H}T\partial^2_{TT}+(2T-4\dot{H})\partial_{T}-1=(4\dot{H}\partial_{T}-1)\hat{M}_{1}=-(\frac{1}{3H}\partial_{t}+1)\hat{M}_{1}
=-\hat{M}_{3}\hat{M}_{1}, 
\end{equation}
\begin{equation}
\hat{M}_3=	\frac{1}{3H}\partial_{t}+1. 
\end{equation}
Using these basic equations we can construct high hierarchy of $F(T)$ gravity. For the case $\rho_m=p_m=0$ such hierarchy can be written as
\begin{equation}
\hat{M}^n_1F_n=0,
\end{equation}
where $F_1=F$. Some equations from this hierarchy for $n=1, 2, 3, ...$ are
\begin{equation}
-2TF_{1T}+F_1=0, 
\end{equation}
\begin{equation}
4T^2F_{2TT}+F_2=0, 
\end{equation}
\begin{equation}
-8T^3F_{3TTT}-12T^2F_{3TT}-2TF_{3T}+F_3=0, 
\end{equation}
and so on. From the system (2.16)-(2.18) follows that any solution of the equation (2.16) automatically solves the equations (2.17)-(2.18). It means 
 that we need just to solve the equation (2.16), as that guarantees  a solution to the equations (2.17) and (2.18).  Finally we present the effective EoS parameter
 \begin{equation}
w_{eff}=-1-3^{-1}H^{-1}[\ln{(\hat{M}_1F)}]_{t}=-1-3^{-1}[\ln{(\hat{M}_1F)}]_{N}.
\end{equation}
\subsection{Particular models of $F(T)$  gravity}
We note that some explicit models of $F(T)$ gravity appeared in the literature (see, e.g. \cite{BF},\cite{L3}, \cite{M6}, \cite{M3}, \cite{Yang4}, \cite{M1}, \cite{WuYu3}). Here we would like to present some new  models of modified teleparallel gravity. 

 \subsubsection{Example 1: The M$_{13}$ - model}
 
 Let us consider the  M$_{13}$ - model. Its   Lagrangian is
 \begin {equation}
F(T)=\sum_{j=-m}^{n}\nu_j(t)T^{j}=\nu_{-m}(t)T^{-m}+ ... +\nu_{-1}(t)T^{-1}+\nu_0(t)+\nu_{1}(t)T + ... + \nu_n(t)T^{n}.
\end{equation}
Consider the particular example when $m=n=1$ and $\nu_j=consts$. Then 
 \begin {equation}
F=\nu_{-1}T^{-1}+\nu_{0}+\nu_{1}T, \quad 
F_{T}=-\nu_{-1}T^{-2}+\nu_{1}, \quad F_{TT}=2\nu_{-1}T^{-3}.
\end{equation}
Substituting these expressions into (2.9)-(2.10) we obtain
\begin{align}
3k^{-2}H^2=\rho_{eff}+ \rho_m, 
\end{align}
\begin{align}
-k^{-2}(2\dot{H}+3H^2)=p_{eff}+p_m, 
\end{align}
where
\begin{align}
\rho_{eff}=k^{-2}[3H^2-1.5\nu_{-1}T^{-1}+0.5\nu_1 T-0.5\nu_0], 
\end{align}
\begin{align}
p_{eff}=k^{-2}[6\nu_{-1}\dot{H}T^{-2}+1.5\nu_{-1} T^{-1}-0.5\nu_1T+0.5\nu_0+2(\nu_1-1)\dot{H}-3H^2]. 
\end{align}
The effective EoS parameter is given by
\begin{align}
w_{eff}=\frac{p_{eff}}{\rho_{eff}}=\frac{6\nu_{-1}\dot{H}T^{-2}+1.5\nu_{-1} T^{-1}-0.5\nu_1T+0.5\nu_0+2(\nu_1-1)\dot{H}-3H^2}{3H^2-1.5\nu_{-1}T^{-1}+0.5\nu_1 T-0.5\nu_0}. 
\end{align}
Let us set $\nu_1=1$. Then
\begin{align}
\rho_{eff}=k^{-2}[-1.5\nu_{-1}T^{-1}-0.5\nu_0], \quad 
p_{eff}=k^{-2}[6\nu_{-1}\dot{H}T^{-2}+1.5\nu_{-1} T^{-1}+0.5\nu_0] 
\end{align}
and 
\begin{align}
w_{eff}=\frac{p_{eff}}{\rho_{eff}}=\frac{6\nu_{-1}\dot{H}T^{-2}+1.5\nu_{-1} T^{-1}+0.5\nu_0}{-1.5\nu_{-1}T^{-1}-0.5\nu_0}=-1-\frac{6\nu_{-1}\dot{H}T^{-2}}{1.5\nu_{-1}T^{-1}+0.5\nu_0}, 
\end{align}
respectively.
\subsubsection{Example 2: The M$_{21}$ - model}
 Our next example is the M$_{21}$ - model
  \begin{align}
F=T+\alpha T^{\delta}\ln{T}. 
\end{align}
Then 
  \begin {equation}
F_{T}=1+\alpha\delta T^{\delta-1}\ln{T}+\alpha T^{\delta-1}, \quad 
F_{TT}=\alpha\delta(\delta-1) T^{\delta-2}\ln{T}+\alpha(2\delta-1) T^{\delta-2}.
\end{equation}
 In this case, Eqs.(2.9)-(2.10) take the form
 \begin{align}
	-T-2\alpha T^{\delta}-\alpha(2\delta-1) T^{\delta}\ln{T}=2k^2 \rho_m, 
\end{align}
\begin{align}
\alpha(2\delta-1)(T-4\delta\dot{H})T^{\delta-1}\ln{T}+T-4\dot{H}+2\alpha T^{\delta}-4\alpha\dot{H}(4\delta-1)T^{\delta-1}=2k^2p_m. 
\end{align}
So we have
\begin{align}
\rho_{eff}=0.5k^{-2}[2\alpha T^{\delta}+\alpha(2\delta-1) T^{\delta}\ln{T}], 
\end{align}
\begin{align}
p_{eff}=-0.5k^{-2}\alpha T^{\delta-1}[(2\delta-1)(T-4\delta\dot{H})\ln{T}+2T-4(4\delta-1)\dot{H}]. 
\end{align}
The special case $\delta=0.5$ deserves separate consideration. In this case the above equations take the more simple form
\begin{align}
	-T-2\alpha T^{0.5}=2k^2 \rho_m, \quad 
	T-4\dot{H}+2\alpha T^{0.5}-4\alpha\dot{H}T^{-0.5}=2k^2p_m. 
\end{align}
For the density of energy and pressure we get the following expressions
\begin{align}
\rho_{eff}=k^{-2}\alpha T^{0.5}, \quad p_{eff}=-k^{-2}\alpha T^{-0.5}(T-2\dot{H}). 
\end{align}
\subsubsection{Example 3: The M$_{22}$ - model}
 Now we consider  the M$_{22}$ - model
  \begin{align}
F=T+f(y), \quad y=\tanh[T]. 
\end{align}
Then 
  \begin {equation}
F_{T}=1+f_y(1-y^2),\quad 
F_{TT}=f_{yy}(1-y^2)^2-2y(1-y^2)f_y
\end{equation}
so that  Eqs.(2.9)-(2.10) take the form
 \begin{align}
	-T-2(1-y^2)Tf_y+f=2k^2 \rho_m, 
\end{align}
\begin{align}
	T-4\dot{H}-8(1-y^2)^2T\dot{H}f_{yy}+(16y\dot{H}T+2T-4\dot{H})(1-y^2)f_{y}-f=2k^2p_m. 
\end{align}
So we have
\begin{align}
\rho_{eff}=0.5k^{-2}[2(1-y^2)Tf_y-f], 
\end{align}
\begin{align}
p_{eff}=0.5k^{-2}[8(1-y^2)^2T\dot{H}f_{yy}-(16y\dot{H}T+2T-4\dot{H})(1-y^2)f_{y}+f]. 
\end{align}
The EoS parameter  reads as
$$
w_{eff}=\frac{8(1-y^2)^2T\dot{H}f_{yy}-(16y\dot{H}T+2T-4\dot{H})(1-y^2)f_{y}+f}{2(1-y^2)Tf_y-f}=
$$
\begin{align}
=-1+\frac{8(1-y^2)^2T\dot{H}f_{yy}-(16y\dot{H}T-4\dot{H})(1-y^2)f_{y}+f}{2(1-y^2)Tf_y-f}. 
\end{align}

\subsubsection{Example 4: The M$_{25}$ - model}
 In this subsubsection  we consider  the M$_{25}$ - model
  \begin{equation}
F=\sum_{-m}^{n}\nu_{j}(t)\xi^{j}, 
\end{equation}
where $\xi=\ln{T}$. As an example we consider the case   $m=n=1, \quad \nu_j=consts$ that is
  \begin {equation}
F=\nu_{-1}\xi^{-1}+\nu_0+\nu_1\xi.
\end{equation}
Then 
  \begin {equation}
F_{\xi}=-\nu_{-1}\xi^{-2}+\nu_1,\quad 
F_{\xi\xi}=2\nu_{-1}\xi^{-3}
\end{equation}
and
  \begin {equation}
F_{T}=(-\nu_{-1}\xi^{-2}+\nu_1)e^{-\xi},\quad 
F_{TT}=(2\nu_{-1}\xi^{-3}+\nu_{-1}\xi^{-2}-\nu_1)e^{-2\xi}.
\end{equation}
For this case, Eqs.(2.9)-(2.10) read as
 \begin{equation}
	2\nu_{-1}\xi^{-2}+\nu_{-1}\xi^{-1}+\nu_0-2\nu_1+\nu_1\xi=2k^2 \rho_m, 
\end{equation}
\begin{equation}
	-4\dot{H}(4\nu_{-1}\xi^{-3}+\nu_{-1}\xi^{-2}-\nu_1)e^{-\xi}-2\nu_{-1}\xi^{-2}-\nu_{-1}\xi^{-1}+2\nu_1-\nu_0-\nu_1\xi=2k^2p_m. 
\end{equation}
	\section{K-essence}

 \subsection{Elements of k-essence}

 The action of k-essence has the form \cite{3}-\cite{32}
\begin {equation}
S=\int d^{4}x\sqrt{-g}[\frac{1}{2\kappa^{2}}R+K(X,\phi)+L_{m}].
\end{equation} 
The corresponding closed set of equations for the FRW metric (2.8) reads as
\begin{equation}
3k^{-2}H^2=2XK_{X}-K+\rho_m, 
\end{equation}
\begin{equation}
	-k^{-2}(2\dot{H}+3H^2)=K+p_m, 
\end{equation}
\begin{equation}
(K_X+2XK_{XX})	\dot{X}+6HXK_X-K_{\phi}=0,
\end{equation}
\begin{equation}
	\dot{\rho_m}+3H(\rho_m+p_m)=0,
\end{equation}
where $X=-0.5\dot{\phi}^2$. The equation of motion of the scalar field $\phi$ is given as
\begin{equation}
	-(a^{3}\dot{\phi}K_{X})_{t}=a^{3}K_{\phi},
\end{equation}
which is just the other form of the equation (3.4). In the purely kinetic k-essence case we have $K_{\phi}=0$ and from the last equation we get (see, e.g. \cite{L1})
\begin{equation}
a^{3}\dot{\phi}K_{X}=a^{3}\sqrt{-2X}K_{X}=\sqrt{\kappa}=const.
\end{equation}

\subsection{Particular models of k-essence}
As examples in this subsection we would like to  present some new types  of k-essence. We believe  that all of them can give rise to cosmic acceleration.
\subsubsection{Example 1: The M$_{12}$ - model}
Let us consider  the   M$_{12}$ - model with the  following Lagrangian 
  \begin {equation}
K=\nu_{-m}(N)N^{-m}+ ... +\nu_{-1}(N)N^{-1}+\nu_0(N)+\nu_1(N)N+ ... + \nu_n(N)N^n,
\end{equation}
where  in general $\nu_j=\nu_j(\phi)=\nu_j(N)$ and $N=\ln{(aa_0^{-1})}$. As an example, we study the case $m=0, n=2, \nu_j=const$. In this case, the M$_{12}$ - model becomes 
\begin{align}
K=\nu_0+\nu_1 N+\nu_2 N^2.
\end{align}
To find $\nu_j$ and $X$ we look for  $H$ for example as
\begin{align}
	H=\mu_0+\mu_1N,
\end{align}
where $\mu_j=consts$ [in general $\mu_j=\mu_j(t)$]. Of course  
\begin{align}
a=a_0e^{N}.
\end{align}
Finally,  we obtain the following   parametric form of the M$_{12}$ - model (parametric purely kinetic k-essence)
\begin{align}
	K=-(2\mu_0\mu_1+3\mu_0^2)-2\mu_1(\mu_1+3\mu_0)N-3\mu_1^2N^2,
\end{align}
\begin{align}
X=k^{-1} a^{6}_0\mu_1^2(\mu_0+\mu_1 N)^2e^{6N}.
\end{align}

\subsubsection{Example 2: The M$_{1}$ - model}
Our next example is   the   M$_{1}$ - model. Its   Lagrangian  looks like
  \begin {equation}
K=\nu_{-m}(t)t^{-m}+...+\nu_{-1}(t)t^{-1}+\nu_0(t)+\nu_1(t)t+ ... +\nu_n(t)t^n,
\end{equation}
where  in general $\nu_j=\nu_j(\phi)=\nu_j(t)$. 
Let us explore  this model for  the case: $m=0, n=2$ and $\nu_j=consts.$ In this case  the M$_1$ - model takes the form 
\begin{align}
K=\nu_0+\nu_1 t+\nu_2 t^2.
\end{align}
To find $\nu_j$ and $X$ we look for  $H$, e.g. as
\begin{align}
	H=\mu_0+\mu_1t
\end{align}
so that 
\begin{align}
a=a_0e^{\mu_0t+0.5\mu_1 t^2},
\end{align}
where $\mu_j=consts$ [in general $\mu_j=\mu_j(t)$]. 
After some calculations  we obtain the following explicit form of the k-essence Lagrangian
\begin{align}
	K=-(2\mu_1+3\mu_0^2)-6\mu_0\mu_1t-3\mu_1^2t^2.
\end{align}
At the same time, we have
\begin{align}
2XK_X=3H^2+K=-2\dot{H}=-2\mu_1.
\end{align}
For $X$ we get the following expression 
\begin{align}
X=\gamma_2^{-1}e^{6\mu_0t+3\mu_1t^2}, \quad \gamma_2^{-1}=\kappa^{-1}a_0^6\mu_1^2.
\end{align}
Hence follows that
\begin{align}
t=\frac{1}{3\mu_1}[-3\mu_0\pm\sqrt{9\mu_0^2+3\mu_1\ln{(\gamma_2X)}}].
\end{align}
Finally, we come to the following M$_{23}$ - model  
\begin{align}
	K=-2\mu_1-3\mu_0^2-\mu_1\ln[\gamma_2X]=\nu_0+\nu_1\ln{X}.
\end{align}
[We recall that in general the M$_{23}$-model reads as
  \begin {equation}
K=\nu_{-m}(t)\zeta^{-m}+...+\nu_{-1}(t)\zeta^{-1}+\nu_0(t)+\nu_1(t)\zeta+ ... +\nu_n(t)\zeta^n,
\end{equation}
where $\zeta=\ln{X}$.]
\subsubsection{Example 3: The M$_{24}$ - model}
Here we present  the  following  M$_{24}$ - model
  \begin {equation}
K=\frac{2m\lambda\sigma^2(-2\beta v+\lambda v^2+\lambda)(1-v^2)}{(\beta-\lambda v)^2}-3[n-\frac{m\lambda\sigma(1-v^2)}{\beta-\lambda v}]^2,
\end{equation} 
\begin{align}
X=\gamma_3(2\beta v-\lambda v^2-\lambda)^2(1-v^2)^2(\beta-\lambda v)^{6m-4},
\end{align}
where $ \gamma_3=\kappa^{-1}\alpha^6m^2\lambda^2\sigma^6$,  $v=\tanh[\sigma t]$ and $\lambda, \sigma, \alpha, \beta, n, m$ are some constants.  Solving the equation (3.3) we obtain
\begin{align}
	H=n-\frac{m\lambda\sigma(1-v^2)}{\beta-\lambda v}
\end{align}
and hence for the scale factor we get the following formula 
\begin{align}
a=\alpha[\beta-\lambda v]^me^{nt}.
\end{align} Note that  
\begin{align}
\dot{H}=\frac{m\lambda\sigma^2(2\beta v-\lambda v^2-\lambda)(1-v^2)}{(\beta-\lambda v)^2}.
\end{align}

\section{Equivalence between $F(T)$-gravity and k-essence}
 In this section, our goal is to study  the relation between modified teleparallel gravity and purely kinetic k-essence. In Appendix C, we will consider this relation in the context with the other modified gravity theories. 
\subsection{General case}
\subsubsection{Variant-I}Consider the  transformation 
\begin {equation}
K=8\dot{H}Tf_{TT}-2(T-2\dot{H})f_{T}+f,
\end{equation}
\begin {equation}
X=\kappa^{-1}k^{-4}a^{6}[\dot{H}+0.5k^2(\rho_m+p_m)]^2,
\end{equation}
where $ T=-6H^2.$ Then Eqs.(2.12)-(2.14) take the form
\begin {equation}
0=-3k^{-2}H^{2}+2XK_X-K+\rho_{m},
\end{equation}
\begin {equation}
0=k^{-2}(2\dot{H}+3H^2)+K+p_m,
\end{equation}
\begin{equation}
(K_X+2XK_{XX})	\dot{X}+6HXK_X=0,
\end{equation}
\begin{equation}
	\dot{\rho_m}+3H(\rho_m+p_m)=0.
\end{equation}
These  are  the equations of motion of purely kinetic k-essence. This result shows that modified teleparallel gravity and purely kinetic k-essence is equivalent to each other, at least in the equation's level. This equivalence allows us to construct a new class of purely kinetic k-essence models starting from some models of  modified teleparallel gravity. 
Let us demonstrate it  for the following modified teleparallel gravity model: $f(T)=\alpha T^n$ \cite{BF}-\cite{L3}. In this case, we have
\begin{align}
	f_T=\alpha n T^{n-1}, \quad f_{TT}=\alpha n(n-1)T^{n-2}.
\end{align}
Substituting these expressions into the equations (4.1)-(4.2) we get
\begin {equation}
K=8\alpha n(n-1)\dot{H}T^{n-1}-2\alpha n(T-2\dot{H})T^{n-1}+\alpha T^n,
\end{equation}
\begin {equation}
X=\kappa^{-1}k^{-4}a^{6}[\dot{H}+0.5k^2(\rho_m+p_m)]^2.
\end{equation}

i) Let $a=a_0e^{g(t)}$ so that $H=\dot{g}, \dot{H}=\ddot{g}$. In this case,  $K$ and $X$ take the form
\begin {equation}
K=8\alpha n(n-1)\ddot{g}(-6)^{n-1}\dot{g}^{2(n-1)}-2\alpha n(-6\dot{g}^2-2\ddot{g})(-6)^{n-1}\dot{g}^{2(n-1)}+\alpha (-6)^n\dot{g}^{2n},
\end{equation}
\begin {equation}
X=\kappa^{-1}k^{-4}a^{6}\ddot{g}^2.
\end{equation}
Now if we consider the simplest case $g=t$ (that means $\dot{g}=1, \ddot{g}=0$),  then we get
\begin {equation}
K=-2\alpha n(-6)^{n}+\alpha (-6)^n=(1-2 n)\alpha (-6)^n,
\end{equation}
\begin {equation}
X=0.
\end{equation}

ii) The more non-trivial model we get, if we consider the example $a=a_0t^m$. In this case $H=mt^{-1}, \dot{H}=-mt^{-2}, T=\frac{-6m^2}{t^2}$ so that  $K$ and $X$ take the form
\begin {equation}
K=8\alpha n(n-1)\dot{H}(\frac{-6m^2}{t^2})^{n-1}-2\alpha n(\frac{-6m^2}{t^2}-2\dot{H})(\frac{-6m^2}{t^2})^{n-1}+\alpha (\frac{-6m^2}{t^2})^n,
\end{equation}
\begin {equation}
X=\kappa^{-1}k^{-4}a_0^{6}m^2t^{6m-4}
\end{equation}
or
\begin {equation}
K=2\alpha m(-6m^2)^{n-1}[-4n(n-1)+2n(1-3m)+3m]t^{-2n},
\end{equation}
\begin {equation}
X=\kappa^{-1}k^{-4}a_0^{6}m^2t^{6m-4}=\gamma_5^{-1}t^{6m-4}.
\end{equation}
Since $t=(\gamma_5 X)^{\frac{1}{6m-4}}$ finally we get the following purely kinetic k-essence model
\begin {equation}
K=2\alpha m(-6m^2)^{n-1}[-4n(n-1)+2n(1-3m)+3m](\gamma_5 X)^{\frac{n}{2-3m}}.
\end{equation}
\subsubsection{Variant-II}
Let us rewrite Eqs.(2.12)-(2.14) as
\begin {equation}
3k^{-2}H^{2}=\rho_{eff}+\rho_{m},
\end{equation}
\begin {equation}
-k^{-2}(2\dot{H}+3H^2=p_{eff}+p_m,
\end{equation}
\begin{equation}
	\dot{\rho}_m+3H(\rho_m+p_m)=0,
\end{equation}
where
\begin {equation}
\rho_{eff}=2T f_{T}-f, \quad p_{eff}=8\dot{H}Tf_{TT}-2(T-2\dot{H})f_{T}+f.
\end{equation}
We now introduce two functions $K$ and $X$ as
\begin {equation}
K=8\dot{H}Tf_{TT}-2(T-2\dot{H})f_{T}+f, \quad X=\frac{4\dot{H}^2(2Tf_{TT}+f_{T})^2}{\kappa a^{-6}}.
\end{equation}
Clearly that these two functions $K$ and $X$ obey the system of the equations (4.3)-(4.6).
\subsection{Special  case: $\phi=\phi_0+\ln{a^{\pm\sqrt{12}}}$}
One of the interesting special cases is when: 
\begin{align}
	\phi=\phi_0+\ln{a^{\pm\sqrt{12}}}.
\end{align}
 It deserves  separate investigation. In fact for this case $\dot{\phi}=\pm\sqrt{12}H$ so that $X=-0.5\dot{\phi}^2=-6H^2=T$. 
 Then the  corresponding continuity equation is
\begin{align}
	\ddot{\phi}(f_T-\dot{\phi}^2f_{TT})+3H\dot{\phi}f_T=0 
\end{align}
or equivalently, in terms of $T$,
\begin{align}
	(f_T+2Tf_{TT})\dot{T}+6HTf_T=0, 
\end{align}
where $\rho^{\prime}=2Tf_T-f, \quad p^{\prime}=f$ and $\dot{\rho}^{\prime}+3H(\rho^{\prime}+p^{\prime})=0$. Now let us split the equation (2.13) into two equations as
\begin{align}
	4\dot{H}Tf_{TT}-(T-2\dot{H})f_{T}=0 
\end{align}
and 
\begin{align}
	-4\dot{H}+T-f=2k^2p_m. 
\end{align}
Eq.(4.27) satisfies automatically since it is just the another form of the continuity equation (4.26). So finally the system of equations of $F(T)$ - gravity
takes the form
\begin{align}
	-T-2Tf_{T}+f=2k^2 \rho_m, 
\end{align}
\begin{align}
	-4\dot{H}+T-f=2k^2p_m, 
\end{align}
\begin{align}
	(f_T+2Tf_{TT})\dot{T}+6HTf_T=0, 
\end{align}
\begin{align}
	\dot{\rho}_m+3H(\rho_m+p_m)=0. 
\end{align}
It    transforms to the  equations (4.3)-(4.6) after the identifications $T=X=-6H^2$ and  $f=2k^{2}K$. So we can conclude that for the special case (4.24) both  $F(T)$ - gravity anf purely kinetic k-essence are equivalent to  each  other at least in the equation's level. Some comments on the continuity equation (4.25) [=(4.26)=(4.27)]. It has two integrals of motion ($I_{jT}=0$):
\begin{align}
I_1=a_0^{-3}a^3T^{0.5}f_T, \quad I_2=	f-a^3T^{0.5}f_T\partial_T^{-1}(a^{-3}T^{-0.5}). 
\end{align}
Its general solution   is given by
\begin{align}
f=C_2+iC_1a_0^2\partial_T^{-1}(a^{-3}T^{-0.5}), \quad C_j=const. 
\end{align}
Finally we would like to present the exact solution both $F(T)$-gravity and purely kinetic k-essence.  As an example let us consider  $\Lambda$CDM for which $a^{-3}=-\frac{1}{2\rho_0}(T+2\Lambda)=-\frac{1}{2\rho_0}(X+2\Lambda)$ so that
\begin{align}
f=f(X)=f(T)= C_2-\frac{iC_1a_0^{3}}{3\rho_0}(T^{1.5}+6\Lambda T^{0.5})= C_2-\frac{iC_1a_0^{3}}{3\rho_0}(X^{1.5}+6\Lambda X^{0.5}), 
\end{align}
which is the M$_{32}$ - model. It is the exact solution of the equations of motion of purely kinetic k-essence and $F(T)$ - gravity simultaneously.

\section{Conclusion}

In this work we investigated the recently developed $F(T)$ gravity, which is a new modified gravity capable of accounting for the present cosmic accelerating expansion with no need of dark energy. $F(T)$ gravity as the modified teleparallel gravity is the extension of the "teleparallel" equivalent of General Gravity (TEGR), which uses the zero curvature Weitzenb$\ddot{o}$ck connection instead of the torsionless Levi-Civita connection, in the same lines as $F(R)$ gravity is the extension of standard General  Gravity. In particular, we presented some new models of $F(T)$ gravity. We analyze the relation between $F(T)$ gravity and k-essence. We also  studied some new models of k-essence namely some parametric models of purely kinetic k-essence. 

It is important to note that $F(T)$ gravity may be consistent with observational data. For instance, the cosmographic parameters of $F(T)$ gravity found in \cite{36} may lie in the region overlapping with those for LCDM. $F(T)$ equation of state [25] or its cosmological perturbations [29]-[30] show very realistic behaviour. $F(T)$ gravity may show LCDM-like, phantom-like or quintessence-like behaviour at dark energy epoch as is shown in our work. Thus, $F(T)$ gravity may pretend to be viable and obserbationally friendly dark energy candidate. New precise observations may select most realistic candidate from existing number of DE models (for recent review see  
\section*{Acknowledgments}
We would like to thank the anonymous referee for providing us with constructive comments and suggestions to improve this work.
\section{Appendix A: Multiple k-essence}

 For the multiple k-essence the action reads as
 \begin {equation}
S=\int d^{4}x\sqrt{-g}[\frac{1}{2\kappa^{2}}R+K(X_1,X_2, ..., X_n, \phi_1, \phi_2, ..., \phi_n)+L_{m}].
\end{equation} 
The corresponding closed set of equations reads as
\begin{align}
3k^{-2}H^2=2\sum_{j=1}^nX_jK_{X_j}-K+\rho_m, 
\end{align}
\begin{equation}
	-k^{-2}(2\dot{H}+3H^2)=K+p_m, 
\end{equation}
\begin{equation}
(K_{X_j}+2X_jK_{X_jX_j})\dot{X_j}+6HX_jK_{X_j}-K_{\phi_j}=0,
\end{equation}
\begin{equation}
	\dot{\rho_m}+3H(\rho_m+p_m)=0,
\end{equation}
where $X_j=-0.5\dot{\phi_j}^2$. The k-essence energy density and pressure respectively, given by
\begin{equation}
	\rho_j=2X_jK_{X_j}-K, \quad p_j=w_j\rho_j.
\end{equation}
The equations of motion of the scalar fields $\phi_j$ are given as
\begin{equation}
-(a^3\dot{\phi}_jK_{X_j})_{t}=a^3K_{\phi_j}.
\end{equation}
Note that the case $K=\sum_{j=1}^nK_j(X_j, \phi_j)$ (mutually non-interacting scalar fields) was investigated in \cite{SD}.

\section{Appendix B: Some models of modified gravity theories and  k-essence}

 In this Appendix B, we present some models of modified gravity theories and  k-essence [$N=\ln{a}, \quad R=6(\dot{H}+2H^2), \quad G=24H^2(\dot{H}+H^2),\quad T=-6H^2,\quad \eta=\int a^{-1}dt=\int H^{-1}a^{-2}da, \quad \xi=\ln{T}, \quad \zeta=\ln{X}, \quad \varsigma=\ln{R}, \quad \vartheta=\ln{G}$]. \\
 1) The M$_1$ - model. Its Lagrangian  has the form
  \begin{equation}
F=K=\sum_{j=-m}^{n}\nu_j(t)t^j.
\end{equation}
2) The M$_2$ - model. The corresponding Lagrangian  reads as
  \begin{equation}
F=K=\sum_{j=-m}^{n}\nu_j(t)e^{jt}.
\end{equation}
3) The M$_3$ - model. It corresponds to the Lagrangian
  \begin{equation}
F=K=\sum_{j=-m}^{n}\nu_j(t)\tanh[jt].
\end{equation}
4) The M$_4$ - model. The Lagrangian of this model is given by
  \begin{equation}
F=K=\sum_{j=-m}^{n}\nu_j(t)\tanh[t]^j.
\end{equation}
5) The M$_5$ - model. The Lagrangian  is given by
  \begin{equation}
F=K=\sum_{j=-m}^{n}\nu_j(t)\cosh[t]^j.
\end{equation}
6) The M$_6$ - model. It reads as
  \begin{equation}
K=\sum_{j=-m}^{n}\nu_j(t)\tan[t]^j.
\end{equation}
7) The M$_7$ - model. It reads as
  \begin{equation}
F=K=\sum_{j=-m}^{n}\nu_j(t)\cos[t]^j.
\end{equation}
8) The M$_8$ - model. It reads as
  \begin{equation}
F=K=\sum_{j=-m}^{n}\nu_j(t)\cosh[jt].
\end{equation}
9) The M$_9$ - model. The corresponding Lagrangian  reads as
  \begin{equation}
K=\sum_{j=-m}^{n}\nu_j(t)t^je^{jt}.
\end{equation}
10) The M$_{10}$ - model. The corresponding Lagrangian  reads as
  \begin{equation}
F=K=\sum_{j=-m}^{n}\nu_j(t)H^{j}.
\end{equation}
11) The M$_{11}$ - model. The corresponding Lagrangian  reads as
\begin{equation}
F=K=\sum_{j=-m}^{n}\nu_j(t)a^{j}.
\end{equation}
12) The M$_{12}$ - model. The corresponding Lagrangian  reads as
\begin{equation}
F=K=\sum_{j=-m}^{n}\nu_j(t)N^{j}.
\end{equation}
13) The M$_{13}$ - model. The corresponding Lagrangian  reads as
\begin{equation}
F=K=\sum_{j=-m}^{n}\nu_j(t)T^{j}.
\end{equation}
14) The M$_{14}$ - model. The corresponding Lagrangian  reads as
\begin{equation}
F=K=\sum_{j=-m}^{n}\nu_j(t)G^{j}.
\end{equation}
15) The M$_{15}$ - model. The corresponding Lagrangian  reads as
\begin{equation}
F=K=\sum_{j=-m}^{n}\nu_j(t)R^{j}.
\end{equation}
16) The M$_{16}$ - model. The corresponding Lagrangian  reads as
\begin{equation}
F=K=\sum_{i,j}\nu_{ij}(t)R^iG^{j}.
\end{equation}
17) The M$_{17}$ - model. The corresponding Lagrangian  reads as
\begin{equation}
F=K=\sum_{i,j}\nu_{ij}(t)R^iT^{j}.
\end{equation}
18) The M$_{18}$ - model. The corresponding Lagrangian  reads as
\begin{equation}
F=K=\sum_{i,j}\nu_{ij}(t)T^iG^{j}.
\end{equation}
19) The M$_{19}$ - model. The corresponding Lagrangian  reads as
\begin{equation}
F=K=\sum_{j=-m}^{n}\nu_j(t)e^{jN}.
\end{equation}
20) The M$_{20}$-model. The corresponding Lagrangian  reads as
\begin{equation}
F=K=\sum_{j=-m}^{n}\nu_j(t)\eta^{j}.
\end{equation}
21) The M$_{21}$-model [see e.g. (2.33)]. \\
22) The M$_{22}$-model [see e.g. (2.41)].\\
23) The M$_{23}$-model (see e.g. (3.23)].\\
24) The M$_{24}$-model [see e.g. (3.24)].\\
25) The M$_{25}$-model [see e.g. (2.48)].\\
26) The M$_{26}$-model. The corresponding Lagrangian  reads as
\begin{equation}
F=K=\sum_{j=-m}^{n}\nu_j(t)\varsigma^{j}.
\end{equation}
27) The M$_{27}$ - model. The corresponding Lagrangian  reads as
\begin{equation}
F=K=\sum_{j=-m}^{n}\nu_j(t)\vartheta^{j}.
\end{equation}
28) The M$_{28}$ - model. The corresponding Lagrangian  reads as
\begin{equation}
F=K=\sum_{j=-m}^{n}\nu_j(t)(\ln{\eta})^{j}.
\end{equation}
29) The M$_{29}$ - model. The corresponding Lagrangian  reads as
\begin{equation}
F=K=\sum_{j=-m}^{n}\nu_j(t)(\tanh[\eta])^{j}.
\end{equation}
30) The M$_{30}$ - model. The corresponding Lagrangian  reads as
\begin{equation}
K=\sum_{j=-m}^{n}\nu_j(t)(\ln[t])^{j}.
\end{equation}
31) The M$_{31}$ - model. The corresponding Lagrangian  reads as
\begin{equation}
K=\sum_{j=-m}^{n}\nu_j(t)(\cosh[R])^{j}.
\end{equation}
32) The M$_{32}$ - model. The corresponding Lagrangian  reads as
\begin{equation}
K=\sum_{j=-m}^{n}\nu_j(t)X^{j}.
\end{equation}

\section{Appendix C: Modified gravity theories as the particular reductions of purely kinetic k-essence}

In this Appendix,  we show that some important modified gravity theories, namely, $F(G), F(R)$ and $F(T)$   can written as  the particular reductions of purely kinetic k-essence. 
\subsection{$F(G)$  gravity}
\subsubsection{Variant-I}Let us consider the following transformation (see, e.g. \cite{O1}- \cite{CST})
\begin {equation}
K=8H^2\ddot{f}_{G}+16H(\dot{H}+H^2)\dot{f}_G+f-Gf_G,
\end{equation}
\begin {equation}
X=\kappa^{-1}k^{-4}a^{6}[\dot{H}+0.5k^2(\rho_m+p_m)]^2.
\end{equation}
Substituting these expressions e.g. into Eqs.(4.3)-(4.6) we get
\begin {equation}
0=-3k^{-2}H^{2}+G f_{G}-f-24\dot{G}H^{3}f_{GG}+\rho_{m},
\end{equation}
\begin {equation}
0=8H^2\ddot{f}_{G}+16H(\dot{H}+H^2)\dot{f}_G+k^{-2}(2\dot{H}+3H^2)+f-Gf_G+p_m,
\end{equation}
\begin{equation}
	\dot{\rho_m}+3H(\rho_m+p_m)=0,
\end{equation}
where
\begin{equation}
 G=24 H^2(\dot{H}+H^2).
\end{equation}
It is the system of the equations of motion of $F(G)$ -gravity with the action 
\begin {equation}
S=\int d^{4}x\sqrt{-g}[\frac{1}{2k^{2}}R+f(G)+L_{m}].
\end{equation}

Now let us consider the particular case when $K=f, \quad X=G$. Then instead of Eqs.(8.3)-(8.5) we obtain the following system
\begin {equation}
0=-3k^{-2}H^{2}+2G f_{G}-f+\rho_{m},
\end{equation}
\begin {equation}
0=k^{-2}(2\dot{H}+3H^2)+f+p_m,
\end{equation}
\begin{equation}
	(f_G+2Gf_{GG})\dot{G}+6HGf_G=0, 
\end{equation}
\begin{equation}
	\dot{\rho_m}+3H(\rho_m+p_m)=0
\end{equation}
and 
\begin{equation}
	Gf_{G}+24\dot{G}H^{3}f_{GG}=0,
\end{equation}
\begin {equation}
0=8H^2\ddot{f}_{G}+16H(\dot{H}+H^2)\dot{f}_G-Gf_G,
\end{equation}
\begin {equation}
\kappa^{-1}k^{-4}a^{6}[\dot{H}+0.5k^2(\rho_m+p_m)]^2=24 H^2(\dot{H}+H^2).
\end{equation}

Let's make two steps back that is let's simplify a problem: 1) we want  reduce the problem to the case $\rho_m=p_m=0$; 2) we want illustrate our results on the pedagogical example: $f(G)=\alpha G^n$. As result, the k-fields take the more simple form
 \begin {equation}
K=\alpha (n-1)G^{n-3}\{8nH^2[(n-2)\dot{G}+G\ddot{G}]+16nH(\dot{H}+H^2)G\dot{G}-G^3\},
\end{equation}
\begin {equation}
X=\kappa^{-1}k^{-4}a^{6}\dot{H}^2,
\end{equation}
\begin {equation}
\phi=\pm i\sqrt{2\kappa^{-1}}k^{-2}\partial_{t}^{-1}(a^{3}\dot{H}).
\end{equation}
Here we want to construct two examples of induced purely kinetic k-essence models: one is in the standard "canonical" form that means in the form $K=K(X)$ and another in the parametric form that means in the form $K=K(t), \quad X=X(t)$ ($t$ plays the role of the parameter). 

i) Let  $a=\beta t^n$. Then the corresponding  purely kinetic k-essence reads as
 \begin {equation}
K=K(X)=16\alpha n^9(n-1)^3[(113-33n)n^3(n-1)(\frac{X}{\gamma})^{\frac{8}{2-3n}}-8(n-2)(\frac{X}{\gamma})^{\frac{11}{4-6n}}].
\end{equation}
Such model we call  the "canonical" k-essence model.  Note that for this case
\begin {equation}
X=\gamma t^{6n-4}, \quad \phi=\phi_0+\frac{i\sqrt{2\gamma}}{3n-1}t^{3n-1}, \quad \gamma=\kappa^{-1}k^{-4}\beta^{6}n^{2}.
\end{equation}

ii) Now we want to present the parametric k-essence model. To do it, let us consider an  example:  $H=\lambda t^m$. In this case the   purely kinetic k-essence equivalent counterpart of the corresponding $F(G)$-model is given by
 \begin {equation}
K=K(t)=24^{n-3}\alpha\lambda^{3(n-3)}(n-1)t^{(3m-1)(n-3)}[m+\lambda t^{m+1}]^{n-3}[K_1+K_2],
\end{equation}
\begin {equation}
X=X(t)=\kappa^{-1}k^{-4}a_0^6n^2\lambda^2t^{2(m-1)}e^{[\frac{6\lambda}{m+1}t^{m+1}]},
\end{equation}
where
$$
K_1=192nm(n-2)\lambda^{5}t^{5m-2}[3m-1+4\lambda t^{m+1}]+
$$
\begin {equation}
+4608nm\lambda^{8}t^{8m-4}[m+\lambda t^{m+1}][(3m-1)(3m-2)+4(4m-1)\lambda t^{m+1}],
\end{equation}
 \begin {equation}
K_2=9216nm\lambda^{8}t^{8m-4}[m+t][m+\lambda t^{m+1}][3m-1+4\lambda t^{m+1}]-13824\lambda^{9}t^{9m-3}[m+\lambda t^{m+1}]^3.
\end{equation}

Such model we call  the parametric k-essence model.
\subsubsection{Variant-II}
We now introduce two functions $K$ and $X$ as
\begin {equation}
K=8H^2\ddot{f}_{G}+16H(\dot{H}+H^2)\dot{f}_G+f-Gf_G, \quad X=\frac{8H^2\ddot{f}_{G}+16H(\dot{H}+H^2)\dot{f}_G-24\dot{G}H^{3}f_{GG}}{4\kappa a^{-6}},
\end{equation}
where $f(G)$ obeys the system (8.3)-(8.5). Then these functions solve the system of the equations of motion of purely kinetic k-essence (4.3)-(4.6).
\subsection{$F(R)$ gravity}
\subsubsection{Variant-I}In this subsection we consider the following  transformation (see, e.g. \cite{O1}- \cite{CST})
\begin {equation}
K=2[\ddot{f}_{R}+2H\dot{f}_{R}+0.5f-(\dot{H}+3H^2)f_{R}],
\end{equation}
\begin {equation}
X=\kappa^{-1}k^{-4}a^{6}[\dot{H}+0.5k^2(\rho_m+p_m)]^2,
\end{equation}
where
\begin{equation}
 R=6(\dot{H}+2H^2).
\end{equation}
The substitution (8.25)-(8.26) into Eqs.(4.3)-(4.6) gives
\begin {equation}
0=-3k^{-2}H^{2}-6H\dot{R}f_{RR}+6(\dot{H}+H^{2})f_{R}-f+\rho_{m},
\end{equation}
\begin {equation}
0=2[\ddot{f}_{R}+2H\dot{f}_{R}+0.5f-(\dot{H}+3H^2)f_{R}]+k^{-2}(2\dot{H}+3H^2)+p_m,
\end{equation}
\begin{equation}
	\dot{\rho_m}+3H(\rho_m+p_m)=0.
\end{equation}
It is the equations of $F(R)$ -gravity. The corresponding  action  is 
\begin {equation}
S=\int d^{4}x\sqrt{-g}[\frac{1}{2k^{2}}R+f(R)+L_{m}].
\end{equation} 

Now let us consider the particular case when $K=f, \quad X=R$. The corresponding continuity equation is $(f_R+2Rf_{RR})\dot{R}+6HRf_R=0$. 
Then instead Eqs.(8.28)-(8.30) we obtain the system
\begin {equation}
0=-3k^{-2}H^{2}+2Rf_R-f+\rho_{m},
\end{equation}
\begin {equation}
0=k^{-2}(2\dot{H}+3H^2)+f+p_m,
\end{equation}
\begin{equation}
	(f_R+2Rf_{RR})\dot{R}+6HRf_R=0, 
\end{equation}
\begin{equation}
	\dot{\rho_m}+3H(\rho_m+p_m)=0
\end{equation}
and
\begin {equation}
0=-6H\dot{R}f_{RR}+6(\dot{H}+H^{2})f_{R}-2Rf_R,
\end{equation}
\begin {equation}
0=\ddot{f}_{R}+2H\dot{f}_{R}-(\dot{H}+3H^2)f_{R},
\end{equation}
\begin {equation}
\kappa^{-1}k^{-4}a^{6}[\dot{H}+0.5k^2(\rho_m+p_m)]^2=6(\dot{H}+2H^2).
\end{equation}
Let us  construct an example of the purely kinetic k-essence model induced by  $F(R)$ - gravity. Let's   simplify a problem: we assume that  $\rho_m=p_m=0$ and $f(R)=\alpha R^n$. As result, the k-essence Lagrangian  takes the  form
 \begin {equation}
K=2\alpha R^{n-3}[n(n-1)(n-2)\dot{R}^2+n(n-1)R\ddot{R}+2n(n-1)HR\dot{R}+0.5R^3-n(\dot{H}+3H^2)R^2].
\end{equation}
Here\begin {equation}
X=\kappa^{-1}k^{-4}a^{6}\dot{H}^2,\quad 
\phi=\pm i\sqrt{2\kappa^{-1}}k^{-2}\partial_{t}^{-1}(a^{3}\dot{H}).
\end{equation}

Now we construct the model for the case  $a=\beta t^l$. Then the corresponding  purely kinetic k-essence reads as
 \begin {equation}
K=K(X)=\varsigma X^{\frac{n}{2-3l}},
\end{equation}
where
\begin {equation}
\varsigma=72\alpha l(2l-1)^{n-1}(6l)^{n-3}[2ln(n-1)(2n-2l-1)+3(2l-1)-nl^2(3l-1)^2]\gamma^{\frac{n}{3l-2}}.
\end{equation}
Note that $X$ and $\phi$ are given by
\begin {equation}
X=\gamma t^{6l-4}, \quad \phi=\phi_0+\frac{i\sqrt{2\gamma}}{3l-1}t^{3l-1}, \quad \gamma=\kappa^{-1}k^{-4}\beta^{6}l^{2}.
\end{equation}
Similarly we can construct a new class k-essence models induced by modified gravity theories. These new k-essence models give the equivalent descriptions of dark energy/matter.
\subsubsection{Variant-II}
If we introduce the following two  functions $K$ and $X$ 
\begin {equation}
K=2[\ddot{f}_{R}+2H\dot{f}_{R}+0.5f-(\dot{H}+3H^2)f_{R}], \quad X=\frac{2[\ddot{f}_{R}+2H\dot{f}_{R}]-6H\dot{R}f_{RR}+4\dot{H}f_{R}}{4\kappa a^{-6}},
\end{equation}
then they satisfy the equations (4.3)-(4.6).

\subsection{$F(R,G)$ gravity}
The action of $F(R,G)$ - gravity is given (see, e.g. \cite{O1}- \cite{CST})
\begin {equation}
S=\int d^{4}x\sqrt{-g}[\frac{1}{2k^{2}}F(R,G)+L_{m}].
\end{equation} 
The corresponding system of equations is given by
\begin {equation}
3k^{-2}H^{2}=\rho_{eff},
\end{equation}
\begin {equation}
-k^{-2}(2\dot{H}+3H^2)=p_{eff},
\end{equation}
\begin{equation}
	\dot{\rho}_{eff}+3H(\rho_{eff}+p_{eff})=0,
\end{equation}
\begin{equation}
	\dot{\rho}_m+3H(\rho_m+p_m)=0.
\end{equation}
Here
\begin{equation}
\rho_{eff}=\frac{1}{F_{R}}\{\rho_m+0.5k^{-2}[RF_R-F-6H\dot{F}_{R}+GF_{G}-24H^3\dot{F}_{G}]\},
\end{equation}
\begin{equation}
p_{eff}=\frac{1}{F_{R}}\{p_m+0.5k^{-2}[-RF_R+F+4H\dot{F}_{R}+2\ddot{F}_{R}-GF_{G}+16H(\dot{H}+H^2)\dot{F}_{G}+8H^2\ddot{F}_{G}]\}.
\end{equation}
Let us consider the following  transformation
\begin {equation}
K=\frac{1}{F_{R}}\{p_m+0.5k^{-2}[-RF_R+F+4H\dot{F}_{R}+2\ddot{F}_{RR}-GF_{G}+16H(\dot{H}+H^2)\dot{F}_{G}+8H^2\ddot{F}_{G}]\},
\end{equation}
\begin {equation}
X=0.25\kappa^{-1}a^{6}F_{R}^{-1}\{\rho_m+p_m+k^{-2}[\ddot{F}_{R}-H\dot{F}_{R}+4H(2\dot{H}-H^2)\dot{F}_{G}+4H^2\ddot{F}_{G}]\}.
\end{equation}
After this transformation, Eqs.(8.46)-(8.49) take the form (4.3)-(4.6).
It is the system of equations of purely kinetic k-essence. So in this sense, both $F(R,G)$ - gravity and purely kinetic k-essence  is equivalent to each other. Hence follow the results of the previous two subsections. In fact $F(G)$ and $F(R)$ are the particular reductions of $F(R,G)$ e.g. as: $F(R)=F(R,0)$ and $F(G)=F(0,G)$.  
\section{Appendix D: Some generalized gas models}
One of interesting class of gas/fluid models is models induced by elliptic functions (see also Refs.\cite{K1}-\cite{K4}). Here  some of such  models [Below,   $\sigma(\rho)$ is  the Weiertrastrass $\sigma(x)$ -
 function,   $\zeta(x)$ is the  Weiertrastrass $\zeta(x)$ -
 function,    $am(x)$ is  the Jacobi amplitude ($am(x)$)  function and so on].
\\
	Table 1. \\
	\begin{equation}
 \begin{array}{|c|c|c|} \hline 
MG-I \quad model &    H=\zeta (t)\\
\hline 
MG-II \quad model &   a=\zeta (t)  \\
\hline 
MG-III \quad model &  H=\sigma(t)\\
\hline 
MG-IV \quad model &  a=\sigma(t)\\
\hline 
MG-V\quad model &  H=\mbox{cn}^{\prime}t\\
\hline 
MG-VI \quad model &  H=\mbox{sn}^{\prime}t\\
\hline 
MG-VII \quad model &H=\mbox{dn}^{\prime}t\\
\hline 
MG-VIII \quad model &  H=\mbox{cn}t  \\
\hline 
MG-IX \quad model &   H=\mbox{sn}t\\
\hline 
MG-X \quad model &  H=\mbox{dn}t\\
\hline
\end{array} \end{equation}
Table 2. 
	\begin{equation}
 \begin{array}{|c|c|}
\hline 
MG-XI \quad model &   p=-B[\zeta(\rho)]^{\alpha}\\
\hline 
MG-XII \quad model &   H=\wp(t)  \\
\hline 
MG-XIII \quad model &  H=\wp(t)^{\prime}\\
\hline 
MG-XIV \quad model &   H=\wp(t)^{\prime\prime}\\
\hline 
MG-XV\quad model &  H=\wp(t)^{\prime\prime\prime}\\
\hline 
MG-XVI \quad model &  H=\wp(t)^{IV}\\
\hline 
MG-XVII \quad model &a\left(t\right)=\wp(t)\\
\hline 
MG-XVIII \quad model & a=\wp(t)^{\prime}  \\
\hline 
MG-XIX \quad model &   a=\wp(t)^{\prime\prime}\\
\hline 
MG-XX \quad model & a=\wp(t)^{\prime\prime\prime}\\
\hline
\end{array} \end{equation}
Table 3. 
	\begin{equation}
 \begin{array}{|c|c|c|} \hline 
MG-XXI \quad model &     p=-B[\wp(\rho)]^{0.5}\\ \hline 

MG-XXII \quad model &  p=-B[\wp(\rho)]^{0.5\alpha} \\ \hline
 
MG-XXIII \quad model &  p=A\rho-B[\wp(\rho)]^{0.5\alpha}\\ \hline 

MG-XXIV \quad model &  p=A\sigma(\rho)-B[\sigma(\rho)]^{-\alpha}\\ \hline 

MG-XXV\quad model &  p=\frac{A}{\zeta(\rho)}-B[\zeta(\rho)]^{\alpha}\\ \hline
 
MG-XXVI \quad model &  p=A[\wp(\rho)]^{-0.5}-B[\wp(\rho)]^{0.5\alpha}\\ \hline 

MG-XXVII \quad model &p=A [am(\rho)]-B[am(\rho)]^{-\alpha}\\ \hline 

MG-XXVIII \quad model & p=A\sigma(\rho)  \\ \hline 

MG-XXIX \quad model  &  p=-B[\sigma(\rho)]^{-\alpha}\\ \hline 

MG-XXX \quad model & p=\frac{A}{\zeta(\rho)}\\ \hline 

MG-XXXI \quad model &p=A [am(\rho)]\\ \hline 

MG-XXXII \quad model &p=-B[am(\rho)]^{-\alpha}\\ \hline
\end{array} 
\end{equation}
\section{Appendix E: Knot Universes from Bianchi type-I models}
Our aim in this Appendix is to present some simplest examples of knot universes for the Bianchi type  - I  model. The   corresponding    metric reads as
  \begin{equation}\label{1}
ds^2=-dt^2+A^2dx_1^2+B^2dx_2^2+C^2dx_3^2,
\end{equation}
where we assume that $t, x_i, A, B, C$ are dimensionless and  $A,B,C$ are functions of $t$ alone. 
For the metric (10.1) it is well-known that the field  equations take the form
\begin{eqnarray}
\frac{\dot{A}\dot{B}}{AB}+\frac{\dot{B}\dot{C}}{BC}+\frac{\dot{C}\dot{A}}{CA}-\rho&=&0,\\
\frac{\ddot{B}}{B}+\frac{\ddot{C}}{C}+\frac{\dot{B}\dot{C}}{BC}+p_1&=&0,\\
\frac{\ddot{C}}{C}+\frac{\ddot{A}}{A}+\frac{\dot{C}\dot{A}}{CA}+p_2&=&0,\\
\frac{\ddot{A}}{A}+\frac{\ddot{B}}{B}+\frac{\dot{A}\dot{B}}{AB}+p_3&=&0,
\end{eqnarray}
where we assume that $p_1\neq p_2\neq p_3$.
Consider some examples.

\subsection{The trefoil knot universe}  Let us consider  the following solution of the system (10.2)-(10.5)
\begin{eqnarray}
A&=&[2+\cos (3t)]\cos (2t), \\
B&=&[2+\cos (3t)]\sin (2t), \\
C&=&\sin (3t)
    \end{eqnarray}
and the corresponding expressions for $\rho$ and $p_i$. The solution (10.6)-(10.8) is the parametric equation of the trefoil knot. For that reason the corresponding universe we call  the trefoil knot universe.  

\subsection{The  figure-eight
knot  universe}  Our second example is given by   the following solution of the system (10.2)-(10.5)
\begin{eqnarray}
A&=&[2+\cos (2t)]\cos (3t), \\
B&=&[2+\cos (2t)]\sin (3t), \\
C&=&\sin (4t).
    \end{eqnarray}and the corresponding expressions for $\rho$ and $p_i$. For this solution the corresponding universe we call  the figure-eight
knot  universe as the solution (10.9)-(10.11) is nothing but the parametric equation of the figure-eight
knot (see also Refs. \cite{Knot1}-\cite{Knot2}). 
   \subsection{Integrable sector}
  Finally, we also can consider the integrable cases of the Bianchi type  - I  model. As an example let us consider the following integrable reduction of the Bianchi type  - I  model: \begin{eqnarray}
iS_{t}+\frac{1}{\omega}[S, W]&=&0,\label{2.1}\\
 iW_{x}+\omega [S, W]&=&0,\label{2.2} \end{eqnarray}
where we introduced the new functions as $A=S_1, B=S_2, C=S_3$ and 
 \begin{equation}\label{2.3}
S=S_i\sigma_i=\begin{pmatrix} S_3 &S^{-}\\S^{+} & -S_3\end{pmatrix}, \quad W=W_i\sigma_i=\begin{pmatrix} W_3 &W^{-}\\W^{+} & -W_3\end{pmatrix}.
 \end{equation}
Here $x$ is for example the cosmological constant that is $x=\Lambda$, $S^2=I, S^{\pm}=S_1\pm iS_2, \quad W^{\pm}=W_1\pm i W_2,$  $[A,B]=AB-BA, $ and $\sigma_i$ are Pauli matrices
\begin{equation}\label{2.3}
\sigma_1=\begin{pmatrix} 0&1\\1& 0\end{pmatrix}, \quad \sigma_2=\begin{pmatrix} 0&i\\-i& 0\end{pmatrix}, \quad \sigma_3=\begin{pmatrix} 1&0\\0& -1\end{pmatrix}.
 \end{equation}
 The  system (10.12)-(10.13) admits the Lax representation of the form
 \begin{equation}
\Phi_{x}=U\Phi,\quad
\Phi_{t}=V\Phi, 
\end{equation} 
where 
 \begin{equation}
U=-\lambda S,\quad V=\frac{i}{\lambda+\omega}W-\frac{i}{\omega}W, \quad W=\begin{pmatrix} W_3&W^{-}\\W^{+}& -W_3\end{pmatrix}.\label{2.2} 
\end{equation} 
The compatable condition of the equations (10.16)
\begin{equation}
U_t-V_x+[U,V]=0\label{CC} 
\end{equation}
gives the system (10.12)-(10.13).
 It is interesting to note that  the system (10.12)-(10.13) is related with the following equations.
 
i) \textit{The M-XCIX equation} \cite{M0}:
 \begin{eqnarray}
iS_{t}+0.25\epsilon_1[S, S_{xx}]+\frac{1}{\omega}[S, W]&=&0,\label{2.1}\\
 iW_{x}+\omega [S, W]&=&0,\label{2.2} \end{eqnarray} 
with the Lax pair $U=-\lambda S,\quad V=\lambda^2V_2+\lambda V_{1}+\frac{i}{\lambda+\omega}W-\frac{i}{\omega}W,$ where $V_2=-i\epsilon_1 S,\quad 
V_1=0.25\epsilon_1[S,S_x]$. 

ii) \textit{The M-LXIV equation} \cite{M0}:\begin{eqnarray}
iS_{t}-\epsilon_2i[ S_{xxx}+6(\alpha^2  S)_{x}]+\frac{1}{\omega}[S, W]&=&0,\label{2.1}\\
 iW_{x}+\omega [S, W]&=&0,\label{2.2} \end{eqnarray}  
 for which the Lax pair looks like $
U=-\lambda S,\quad V=\lambda^3V_3+\lambda^2V_2+\lambda V_{1}+\frac{i}{\lambda+\omega}W-\frac{i}{\omega}W$ and where
$V_3=4i\epsilon_2 S,\quad V_2=-\epsilon_2[S,S_x],\quad V_1=-\epsilon_2i( S_{xx}+6\alpha^2  S)$.

ii) \textit{The M-XCIV equation} \cite{M0}:
\begin{eqnarray}
iS_{t}+0.25\epsilon_1[S, S_{xx}]-\epsilon_2i[ S_{xxx}+6(\alpha^2  S)_{x}]+\frac{1}{\omega}[S, W]&=&0,\label{2.1}\\
 iW_{x}+\omega [S, W]&=&0,\label{2.2} \end{eqnarray} 
 with the Lax pair 
 \begin{eqnarray}
U&=&-i\lambda S,\label{2.1}\\
V&=&\lambda^3V_3+\lambda^2V_2+\lambda V_{1}+\frac{i}{\lambda+\omega}V_{-1}-\frac{i}{\omega}V_{-1},\label{2.2} 
\end{eqnarray} 
where \cite{M0}
\begin{eqnarray}
V_3&=&4i\epsilon_2 S,\label{2.8}\\
V_2&=&-i\epsilon_1 S-\epsilon_2[S,S_x],\label{2.8}\\
V_1&=&0.25\epsilon_1[S,S_x]-\epsilon_2i( S_{xx}+6\alpha^2  S),\label{2.1}\\
V_{-1}&=&W=\begin{pmatrix} W_3&W^{-}\\W^{+}& -W_3\end{pmatrix}.\label{2.2} 
\end{eqnarray} 
 \section{ Appendix J: Some generalizations of $F(T)$ gravity}
 In this section we present 3 generalizations of the Friedmann equations of $F(T)$ gravity. Recall that the modified Friedmann equations of the  loop quantum cosmology  (LQC) read as
 \begin{eqnarray} \label{fe1}
3H^2&=&8\pi G\rho\left(1-\frac{\rho}{\rho_c}\right),\\
\dot{H}&=&-4\pi G(\rho+p)\left(1-\frac{2\rho}{\rho_c}\right),\\
\dot{\rho}&=&-3H(\rho+p).
\end{eqnarray}  
\subsection{The M$_{35}$ - model} The M$_{35}$ - model is the given by
\begin{eqnarray} \label{fe1}
3H^2&=&8\pi G\rho\sqrt{1-\frac{2\rho}{\rho_c}},\\
\dot{H}&=&-\frac{4\pi G(\rho+p)}{\sqrt{1-\frac{2\rho}{\rho_c}}}\left(1-\frac{3\rho}{\rho_c}\right),\\
\dot{\rho}&=&-3H(\rho+p).
\end{eqnarray}

\subsection{The M$_{36}$ - model} For the FRW metric the equations of the M$_{36}$ - model read as
\begin{eqnarray} \label{fe1}
3H^2&=&8\pi G\rho_c\sqrt{1-\frac{2\rho}{\rho_c}}\Big(1-\sqrt{1-\frac{2\rho}{\rho_c}}\Big),\\
\dot{H}&=&-8\pi G(\rho+p)\left(1-\frac{1}{2\sqrt{1-\frac{2\rho}{\rho_c}}}\right),\\
\dot{\rho}&=&-3H(\rho+p).
\end{eqnarray}
In the case $2\rho<\rho_c$ we have $\sqrt{1-\frac{2\rho}{\rho_c}}\approx 1-\frac{\rho}{\rho_c}$ so that the previous two systems (11.4)-(11.6) and (11.7)-(11.9) tranforms to the usual system of equations of LQC (11.1)-(11.3).
\subsection{The M$_{37}$ - model}
 
Let us consider the M$_{37}$ - model. Its action is
 \begin {equation}
 {\cal S}=\int d^4 xe[F(R,T)+L_m],
 \end{equation}
 where $R$ is the "curvature" scalar  and $T$ is the "torsion" scalar.  For simplicity in this paper we work in such FRW spacetime in which $R$ and $T$ are given by
 \begin{eqnarray} \label{fe1}
R&=&u+6\epsilon_1(\dot{H}+2H^2),\\
T&=&v+6\epsilon_2H^2,
\end{eqnarray}
where in general  $u=u(t, a,\dot{a}, \ddot{a},\dddot{a}, ...; f_i)$ and $v=v(t,a,\dot{a}, \ddot{a},\dddot{a}, ...; g_i)$ are some real functions, $H=(\ln a)_t$,  $f_i$ and $g_i$ are some unknown functions related with the geometry of the spacetime. Here  we restrict ourselves to the case when  $u=u(a,\dot{a})$ and $v=v(a,\dot{a})$. Then the FRW   equations of the M$_{37}$ - model look like
 \begin{eqnarray}
 D_2F_{RR}+D_1F_R+JF_{RT}+E_1F_T+KF&=&-2a^3\rho,\,\,\,\\
  U+C_2F_{RRT}+C_1F_{RTT}+C_0F_{RT}+MF  &=&6a^2p,\,\,\,\\
 \dot{\rho}-3H(\rho+p)&=&0.
 \end{eqnarray} 
 Here
 \begin{eqnarray} \label{3.10}
 D_2&=&-6\epsilon_1\dot{R}a^2\dot{a},\\ \label{3.11}
   D_1&=&6\epsilon_1a^2 \ddot{a}+a^3u_{\dot{a}}\dot{a},\\  \label{3.12}
  J&=&-6\epsilon_1a^2 \dot{a}\dot{T},\\ \label{3.13}
 E_1&=&12\epsilon_2a \dot{a}^2+a^3v_{\dot{a}}\dot{a},\\ \label{3.14}
 K&=&-a^3
 \end{eqnarray}
and
\begin{eqnarray}
 U&=&A_3F_{RRR}+A_2F_{RR}+A_1F_{R},\label{3.15}\\
 A_3&=&-6\epsilon_1\dot{R}^2a^2,\label{3.16}\\
A_2&=&-12\epsilon_1\dot{R}a\dot{a}-6\epsilon_1\ddot{R}a^2+a^3\dot{R}u_{\dot{a}},\label{3.17}\\
   A_1&=&12\epsilon_1\dot{a}^2+6\epsilon_1a \ddot{a}+3a^2\dot{a}u_{\dot{a}}+a^3\dot{u}_{\dot{a}}-a^3u_a,\label{3.18}\\
 B_2&=&12\epsilon_2\dot{T}a \dot{a}+a^3\dot{T}v_{\dot{a}},\label{3.19}\\
 B_1&=&24\epsilon_2\dot{a}^2+12\epsilon_2a \ddot{a}+3a^2\dot{a}v_{\dot{a}}+a^3\dot{v}_{\dot{a}}-a^3v_a,\label{3.20}\\
 C_2&=&-12\epsilon_1a^2\dot{R}\dot{T},\label{3.21}\\
 C_1&=&-6\epsilon_1a^2\dot{T}^2,\label{3.22}\\
C_0&=&-12\epsilon_1\dot{T}a\dot{a}+12\epsilon_2\dot{R}a\dot{a}-6\epsilon_1a^2\ddot{T}+a^3\dot{R}v_{\dot{a}}+a^3\dot{T}u_{\dot{a}},\label{3.23}\\
 M&=&-3a^2. \label{3.24}
 \end{eqnarray}

 It is interesting to note that the M$_{37}$ - model (\ref{3.1}) admits some interesting particular
  and physically important  cases. Some particular cases are now presented.

 \textit{i) The M$_{44}$ - model.}
 Let the function $F(R,T)$ be independent from the torsion scalar $T$ that is $F=F(R,T)=F(R)$.
  Then the  action (\ref{3.1}) acquires the form
   \begin{equation}\label{3.25}
 S_{44}=\int d^4 x e[F(R)+L_m],
 \end{equation}
 where
 \begin{equation}\label{3.26}
 R=u+R_s=u+\epsilon_1g^{\mu\nu}R_{\mu\nu},
 \end{equation}
  is the curvature scalar. It is the M$_{44}$ - model. We work with  the FRW metric.
   In this case  $R$ takes  the form
 \begin{equation} \label{3.27}
R=u+6\epsilon_1(\dot{H}+2H^2).
\end{equation}
The action can be rewritten as
 \begin{equation}\label{3.28}
 S_{44}=\int dtL_{44},
 \end{equation}
 where the   Lagrangian is given by
 \begin{equation}\label{3.29}
  L_{44}=a^3[F-(R-u)F_R+L_m]-
 6\epsilon_1F_Ra\dot{a}^2-6\epsilon_1F_{RR}\dot{R}a^2\dot{a}.
 \end{equation}
 The corresponding field equations of the M$_{44}$ - model read as
 \begin{eqnarray}
 D_2F_{RR}+D_1F_R+KF&=&-2a^3\rho,\nonumber\\
   A_3F_{RRR}+A_2F_{RR}+A_1F_{R}+MF  &=&6a^2p,\label{3.30}\\
 \dot{\rho}+3H(\rho+p)&=&0.\nonumber
 \end{eqnarray}
 Here
 \begin{eqnarray} \label{3.31}
 D_2&=&-6\epsilon_1\dot{R}a^2\dot{a},\\ \label{3.32}
   D_1&=&6\epsilon_1a^2 \ddot{a}+a^3u_{\dot{a}}\dot{a},\\
   K&=&-a^3 \label{3.33}
 \end{eqnarray}
and
\begin{eqnarray}
  A_3&=&-6\epsilon_1\dot{R}^2a^2,\label{3.34}\\
A_2&=&-12\epsilon_1\dot{R}a\dot{a}-6\epsilon_1\ddot{R}a^2+a^3\dot{R}u_{\dot{a}},\label{3.35}\\
   A_1&=&12\epsilon_1\dot{a}^2+6\epsilon_1a \ddot{a}+3a^2\dot{a}u_{\dot{a}}+a^3\dot{u}_{\dot{a}}-a^3u_a,\label{3.36}\\
  M&=&-3a^2. \label{3.37}
 \end{eqnarray}
 If $u=0$ then we get the following equations of the  standard  $F(R_s)$ gravity (after $R=R_s$):
 \begin{eqnarray}
 6\dot{R}HF_{RR}-(R-6H^2)F_R+F&=&\rho,\label{3.38}\\
   -2\dot{R}^2F_{RRR}+[-4\dot{R}H-2\ddot{R}]F_{RR}+[-2H^2-4a^{-1} \ddot{a}+R]F_{R}-F &=&p,\label{3.39}\\
 \dot{\rho}+3H(\rho+p)&=&0.\label{3.40}
 \end{eqnarray}

 \textit{ii) The M$_{45}$ - model.}
The action of the M$_{45}$ - model looks like
 \begin{equation}\label{3.41}
 S_{45}=\int d^4 xe [F(T)+L_m],
 \end{equation}
 where $e={\rm det}\,(e_\mu^i)=\sqrt{-g}$ and the torsion scalar $T$ is defined as
 \begin{equation}\label{3.42}
 T=v+T_s=v+\epsilon_2{S_\rho}^{\mu\nu}\,{T^\rho}_{\mu\nu}.
 \end{equation}
 Here
 \begin{eqnarray}
 {T^\rho}_{\mu\nu} &\equiv &-e^\rho_i\left(\partial_\mu e^i_\nu
 -\partial_\nu e^i_\mu\right),\label{3.43}\\
 {K^{\mu\nu}}_\rho &\equiv &-\frac{1}{2}\left({T^{\mu\nu}}_\rho
 -{T^{\nu\mu}}_\rho-{T_\rho}^{\mu\nu}\right), \label{3.44}\\
 {S_\rho}^{\mu\nu} &\equiv &\frac{1}{2}\left({K^{\mu\nu}}_\rho
 +\delta^\mu_\rho {T^{\theta\nu}}_\theta-
 \delta^\nu_\rho {T^{\theta\mu}}_\theta\right).\label{3.45}
 \end{eqnarray}
 For a spatially flat FRW metric \eqref{2.18},  we have the torsion scalar  in the form
 \begin{equation}\label{3.46}
 T=v+T_s=v+6\epsilon_2H^2.
 \end{equation}
 The action \eqref{3.41} can be written as
 \begin{equation}\label{3.47}
 S_{45}=\int dt L_{45},
 \end{equation}
where
 the point-like Lagrangian reads
 \begin{equation}\label{3.48}
  L_{45}=a^3[F-(T-v)F_T+L_m]+6\epsilon_2F_Ta\dot{a}^2.
 \end{equation}
  So finally we get the following  equations of the M$_{45}$ - model:
  \begin{eqnarray}
 E_1F_T+KF&=&-2a^3\rho,\nonumber\\
   B_2F_{TT}+B_1F_{T}+MF  &=&6a^2p,\label{3.49}\\
 \dot{\rho}+3H(\rho+p)&=&0.\nonumber
 \end{eqnarray}
 Here
 \begin{eqnarray} \label{3.50}
  E_1&=&12\epsilon_2a \dot{a}^2+a^3v_{\dot{a}}\dot{a},\\
 K&=&-a^3  \label{3.51}
 \end{eqnarray}
and
\begin{eqnarray} \label{3.52}
  B_2&=&12\epsilon_2\dot{T}a \dot{a}+a^3\dot{T}v_{\dot{a}},\\
 B_1&=&24\epsilon_2\dot{a}^2+12\epsilon_2a     \ddot{a}+3a^2\dot{a}v_{\dot{a}}+a^3\dot{v}_{\dot{a}}-a^3v_a, \label{3.53} \\     M&=&-3a^2.       \label{3.54}
 \end{eqnarray}
 If we put $v=0$ then the M$_{45}$ - model reduces to the usual $F(T_s)$ gravity, where
 $T_s=6\epsilon_2H^2$. As is well-known the equations of $F(T_s)$ gravity are given by
 \begin{eqnarray}
 12H^2 F_T+F&=&\rho,\label{3.55}\\
 48H^2 F_{TT}\dot{H}-F_T\left(12H^2+4\dot{H}\right)-F
 &=&p,\label{3.56}\\
 \dot{\rho}+3H(\rho+p)&=&0,\label{3.57}
 \end{eqnarray}
 where we must put $T=T_s$. Finally we note that it is well-known that the standard
 $F(T_s)$ gravity is not local Lorentz invariant.
  In this context, we have a very meager hope that the M$_{45}$ - model
  \eqref{3.41}  is free from such problems.

\end{document}